\documentclass[preprints,article,accept,pdftex,moreauthors]{Definitions/mdpi} 
\DeclareGraphicsExtensions{.pdf,.prb,.jpg}
\usepackage[percent]{overpic}

\usepackage[version=4]{mhchem}

\newcommand{\JPB}[1]{{#1}}

\graphicspath{{./Figures/}}

\firstpage{1} 
\pubvolume{1}
\issuenum{1}
\articlenumber{0}
\pubyear{2025}
\copyrightyear{2025}
\datereceived{ } 
\daterevised{ } 
\dateaccepted{ } 
\datepublished{ } 
\hreflink{https://doi.org/} 

    \Title{{Quantum} 
 Interference Supernodes, Thermoelectric Enhancement, and the Role of Dephasing}

	\TitleCitation{{Quantum} 
 Interference Supernodes, Thermoelectric Enhancement, and the Role of Dephasing}

	\Author{{Justin} 
 P. Bergfield \orcidA{}} 
	
\AuthorNames{Justin P. Bergfield} 
\isAPAStyle{%
\AuthorCitation{Bergfield, J.P.} 
}{%
\isChicagoStyle{%
\AuthorCitation{Justin P. Bergfield} 
}{
\AuthorCitation{Bergfield, J.P.} 
}
}

\address[1]{{Department} 
 of Physics, Illinois State University, Normal, IL 61761, USA; jpbergf@ilstu.edu}

\abstract{
Quantum interference 
can strongly enhance thermoelectric response, with higher-order ``supernodes'' predicted to yield scalable gains in thermopower and efficiency.  
A central question, however, is whether such features are intrinsically more fragile to dephasing.  
Using B\"uttiker voltage--temperature probes, we establish an \emph{{order-selection rule}
}: the effective near-node order is set by the lowest among coherent and probe-assisted channels.  
Supernodes are therefore fragile in an absolute sense because their transmission is parametrically suppressed with order.  
However, once an incoherent floor dominates, the fractional suppression of thermopower, efficiency, and figure of merit becomes universal and order-independent. 
Illustrating these principles with benzene- and biphenyl-based junction calculations, we show that the \emph{{geometry}} 
of environmental coupling---through a single orbital or across many---dictates whether coherence is lost by order reduction or by floor building.  
These results yield general scaling rules for the thermoelectric response of interference nodes under dephasing.
}


\begin{document}

\section{Introduction }

Quantum interference (QI) is among the most striking hallmarks of phase-coherent transport at the nanoscale.  
Owing to the dual wave- and particle-like nature of quantum excitations, electronic conduction need not follow any classical path: coherent amplitudes can superpose destructively, completely suppressing current flow and producing transmission zeros (nodes) in the electronic transmission function $\mathcal{T}(E)$.  
Such nodes strongly influence both charge transport and the thermoelectric response%
~\cite{evers2020advances,dubi2011colloquium, miroshnichenko2010fano,solomon2008quantum,solomon2011small,cardamone2006controlling,stafford2007quantum,lambert2015basic,bergfield2009many,bergfield2010coherent,bergfield2010giant,bergfield2014thermoelectric,bergfield2011novel}.  
Their positions and lineshapes are dictated by the symmetries of the full Hamiltonian, including geometric, topological, and~many-body aspects, so nodes serve not only as fingerprints of microscopic symmetry but also as potential resources for quantum-enhanced functionality~\cite{basov2017towards,evers2020advances,dubi2011colloquium,cardamone2006controlling}.

Thermoelectric performance is particularly sensitive to QI, with~a dependence that can generally be related to how rapidly $\mathcal{T}(E)$ varies with energy. 
Near a node, charge and entropy currents are suppressed in different ways, and~their ratio, the~thermopower, 
is predicted to exhibit strong enhancement%
~\cite{bergfield2009thermoelectric,bergfield2010giant}. 
{Related mechanisms have also been demonstrated in superconducting nanostructures, where spin-splitting, phase coherence, and~order–parameter symmetry yield unusually large thermoelectric responses and motivate novel device concepts%
~\cite{ozaeta2014predicted,kolenda2016observation,kalenkov2017large,marchegiani2020nonlinear,guarcello2023thermoelectric,guarcello2023bipolar,germanese2022bipolar}.} 
The \emph{{order}
} of a node, defined by the local scaling 
$\mathcal{T}(E) \propto (E-E_0)^{2n}$ near $E_0$, controls the magnitude and scaling of this enhancement%
~\cite{bergfield2010giant,bergfield2010coherent,wierzbicki2011influence}. 
Large violations of the Wiedemann–Franz law are expected in these regimes, reflecting a breakdown of the 
free-electron picture and offering opportunities for quantum-engineered thermoelectricity%
~\cite{bergfield2009thermoelectric,bergfield2010giant,majidi2024heat,bell2008cooling,disalvo1999thermoelectric,shi2020advanced,rincon2016thermopower}.

In systems composed of $N$ node-bearing subunits, destructive interference {\em {can}} 
combine to produce a higher-order {\em {supernode}}, 
in~which coincident quadratic nodes yield a $2N$-th order suppression of transport~\cite{bergfield2010giant,barr2013transmission}.
Such supernodes are predicted within single-determinant, effective single-particle theories, e.g.,~extended H\"uckel, Hartree--Fock, or~standard Kohn--Sham DFT, where the response function factorizes across connected subunits and local zeros combine in a ``series-propagation’’ manner~\cite{solomon2011small}.  
By contrast, a~full many-body treatment reveals a different structure~\cite{solomon2011small,barr2013transmission,pedersen2014quantum}.    
Because the interaction self-energy is intrinsically nonlocal, this simple factorization generally fails.  
As a result, pure supernodes are likely difficult to realize experimentally: unless protected by exact symmetries, they tend to fragment into multiple quadratic nodes or be lifted altogether~\cite{barr2013transmission,pedersen2014quantum}.
Even so, sharp interference features with effective order $n>1$ are neither rare nor irrelevant, e.g.,~arising as Fano antiresonances, quasi-bound states adjacent to a node, or~clusters of nearby zeros that can mimic higher-order behavior over experimentally relevant windows~\cite{miroshnichenko2010fano,bergfield2010giant,solomon2011small}.
Because thermoelectric and thermodynamic responses scale sensitively with effective node order, understanding how dephasing reshapes such higher-order structures is~essential.

Decoherence and dephasing occur when coherent electron flow couples to external degrees of freedom, e.g.,~vibrations, solvent, fluctuating charges, or~other environmental baths.  
Such couplings are unavoidable, and~their influence on interference features is therefore central to any realistic description of device performance.  
We model dephasing using the probe concept introduced by B\"uttiker~\cite{buttiker1986four,buttiker1988coherent}, in which fictitious terminals absorb and re-emit carriers stochastically, randomizing their phase while conserving macroscopic currents~\cite{kilgour2016inelastic,kilgour2015charge,golizadeh2007nonequilibrium}. %
Two variants are commonly employed: a voltage probe (VP), which enforces local charge conservation ($I_P^{(0)}=0$), and~a voltage--temperature probe (VTP), which enforces both charge and heat conservation ($I_P^{(0)}=I_P^{(1)}=0$) \cite{engquist1981definition,bergfield2013probing,meair2014local,Stafford2016local,Shastry2015local}.  
Although sometimes treated as interchangeable, the~two impose distinct thermodynamic conditions~\cite{eregonarXiv}.  
At finite temperature the distinction is crucial: a VP does not enforce local equilibrium and can act as an entropy source or sink, whereas a VTP imposes full local thermodynamic equilibrium.  
In this work, we therefore focus on VTPs, reverting to VPs only in cases where their predictions~coincide.


Here we investigate how dephasing modifies the thermoelectric and thermodynamic response of systems with transmission supernodes.  
Using single and multiple VTPs to model decoherence, and~exploiting the near-node universality of the transmission~\cite{bennett2024quantum}, we establish a simple \emph{{order-selection rule}} 
to determine the effective node order in the presence of dephasing.  
Supernodes are therefore more fragile in an absolute sense, since their transmission is parametrically weaker with increasing order.  
However, once an incoherent floor develops, the~\emph{{fractional}} 
suppression of thermopower, efficiency, and~figure of merit becomes universal and effectively order-independent.  
We further show that the scaling of the response depends not only on node order but also on the geometry of probe--orbital coupling, suggesting strategies to either mitigate or exploit dephasing in practical~applications.

\section{Quantum Transport~Theory}

We investigate how node order and dephasing shape charge and heat transport in interacting open quantum systems, modeled as a nanosystem coupled to \(M\) macroscopic electrodes treated as ideal Fermi gases. Transport is analyzed within the non-equilibrium Green’s function (NEGF) formalism, which provides a rigorous framework for such systems~\cite{HaugAndJauhoBook, stefanucci2013nonequilibrium, cuevas2010molecular}. Our focus is on systems whose transport is predominantly elastic \linebreak  and~phase-coherent.  

In the linear-response regime, the~steady-state current of order \(\nu\) flowing into electrode \(\alpha\) {is} 
\begin{equation}
	\label{eq:linear_response}
	I_\alpha^{(\nu)} = \sum_{\beta=1}^{M} \left[ {\cal L}^{(\nu)}_{\alpha\beta} (\mu_\beta - \mu_\alpha) + {\cal L}^{(\nu+1)}_{\alpha\beta} \frac{T_\beta - T_\alpha}{T_0} \right],
\end{equation}
where \(\nu=0\) corresponds to particle current and \(\nu=1\) to heat current.  
The Onsager coefficients are
\begin{equation}
	{\cal L}^{(\nu)}_{\alpha\beta} = \frac{1}{h} \int dE\, (E - \mu_0)^\nu\, {\cal T}_{\alpha\beta}(E) \left(-\frac{\partial f_0}{\partial E} \right),
\end{equation}
with \(f_0(E) = [ \exp((E - \mu_0)/k_B T_0) + 1 ]^{-1}\) being the equilibrium Fermi--Dirac distribution at chemical potential \(\mu_0\) and temperature \(T_0\).  
In coherent transport, the~transmission function between leads \(\alpha\) and \(\beta\) {is} 
\begin{equation}
	\label{eq:transmission_prob}
	\mathcal{T}_{\alpha\beta}(E) = {\rm Tr}\!\left[ \mathbf{\Gamma}_\alpha(E)\, \mathbf{\mathcal{G}}(E)\, \mathbf{\Gamma}_\beta(E)\, \mathbf{\mathcal{G}}^\dagger(E) \right],
\end{equation}
where \(\mathbf{\mathcal{G}}(E)\) is the junction Green’s function and \(\mathbf{\Gamma}_\alpha(E)\) is the tunneling-width matrix for lead \(\alpha\), defined as
\begin{equation}
	[\mathbf{\Gamma}_\alpha(E)]_{nm} = 2\pi \sum_{k\in\alpha} V_{nk} V_{mk}^\ast\, \delta(E-\epsilon_k),
\end{equation}
with \(n,m\) as molecular orbital indices, and \(V_{nk}\) the coupling between orbital \(n\) and electrode state \(k\) of energy \(\epsilon_k\).  
In the wide-band limit considered here, \(\mathbf{\Gamma}_\alpha\) is~energy-independent.  

Within molecular Dyson equation (MDE) theory~\cite{bergfield2009many}, the junction Green’s function can be written exactly as
\begin{equation}
	\label{eq:Dyson2}
	\mathbf{\mathcal{G}}(E) = \left[ \mathbf{\mathcal{G}}_{\rm mol}^{-1}(E) - \mathbf{\Sigma}_{\rm T}(E) - \Delta\mathbf{\Sigma}_{\rm C}(E) \right]^{-1},
\end{equation}
where \(\mathbf{\mathcal{G}}_{\rm mol}(E)\) is the Green’s function of the isolated molecule, \(\mathbf{\Sigma}_{\rm T}(E) = -\tfrac{i}{2} \sum_{\alpha=1}^{M} \mathbf{\Gamma}_\alpha\) is the total tunneling self-energy, and~\(\Delta\mathbf{\Sigma}_{\rm C}(E)\) is the Coulomb self-energy correction due to resonance broadening.  
In the elastic cotunneling regime considered here, \(\Delta\mathbf{\Sigma}_{\rm C} \approx 0\) and inelastic contributions are negligible~\cite{bergfield2009many}. 
Finally, the~molecular Green’s function admits the Lehmann representation~\cite{bergfield2009many}
\begin{equation}
	\label{eq:Gmol}
	[\mathbf{\mathcal{G}}_{\rm mol}(E)]_{n\sigma,m\sigma'} = \sum_{\nu,\nu'} \frac{{\cal P}(\nu) + {\cal P}(\nu')}{E - E_{\nu'} + E_\nu + i0^+} \langle \nu | d_{n\sigma} | \nu' \rangle \langle \nu' | d^\dagger_{m\sigma'} | \nu \rangle,
\end{equation}
where \(d^\dagger_{n\sigma}\) (\(d_{n\sigma}\)) creates (annihilates) an electron of spin \(\sigma\) on orbital \(n\), \(|\nu\rangle\) is an eigenstate of the molecular Hamiltonian with energy \(E_\nu\), and~\({\cal P}(\nu)\) is its grand-canonical occupation probability at \((\mu_0,T_0)\).

\subsection*{{Model} 
 Hamiltonian and~Parameters}
\label{sec:hamiltonian}


To illustrate our results, 
we consider two representative single-molecule junctions (SMJs) whose $\pi$-conjugated backbones capture the essential physics of interference nodes: Au-1,3-benzenedithiol-Au (BDT) in the meta configuration, which exhibits a quadratic node, and~
Au-3,3$'$-biphenyldithiol-Au (BPDT), where connectivity and torsion-controlled inter-ring coupling give rise to effective higher-order interference features~\cite{markussen2010relation,solomon2008understanding,liu2018quantum}.  
\JPB{We focus on these junctions because they are chemically stable, experimentally benchmarked molecules that exhibit canonical interference features, providing minimal yet physically realistic models for analyzing how dephasing reshapes transport in the vicinity of supernodes.}

The effective Hamiltonian for the $\pi$-system can be derived from first principles by integrating out 
off-resonant degrees of freedom (e.g., the~$\sigma$-system, image-charge effects, and~substrate polarization), 
which are absorbed into renormalized site energies and couplings~\cite{barr2012effective}. 
In a localized orbital basis, the~resulting one-body Hamiltonian is
\begin{equation}
	\label{eq:Hmol}
	H_{\rm mol} = \sum_{n} \varepsilon_{n} \hat{\rho}_{n} 
	- \sum_{\langle n,m\rangle} t_{nm} \hat{d}^\dagger_{n} \hat{d}_{m},
\end{equation}
where $\varepsilon_{n}$ is the effective on-site potential, $t_{nm}$ is the effective hopping matrix element, 
$\hat{\rho}_{n} = \hat{d}^\dagger_{n} \hat{d}_{n}$ is the local charge density, and~$\langle n,m \rangle$ denotes 
nearest-neighbor pairs with $n \neq m$.  
\JPB{In this notation, $t_\perp$ denotes the inter-ring hopping in biphenyl and is simply one of the $t_{nm}$ values connecting the two phenyl subunits.}
Equation~\eqref{eq:Hmol} is formally equivalent to an extended H\"uckel Hamiltonian in which electron--electron 
interactions are neglected. The~influence of interactions on thermoelectric transport has been analyzed in detail elsewhere~\cite{barr2012effective,barr2013transport}.  \JPB{Couplings to electrodes and probes are included through their self-energies within the NEGF formalism.}

For the BDT junction we take the electrode couplings to be symmetric, $\Gamma_L = \Gamma_R \approx 0.44~{\rm eV}$, 
with nearest-neighbor hopping $t_{nm} = 2.6~{\rm eV}$, values obtained from fits to experiment~\cite{barr2012effective,bergfield2009many}.
For the BPDT junction, comparison with measured thermopower 
$S = (12.9 \pm 2.2)\,\upmu{\rm V/K}$ and conductance 
$G = (1.7 \pm 0.2)\times 10^{-4} G_0$ yields $\Gamma_L = \Gamma_R \approx 0.21~{\rm eV}$ \cite{mishchenko2009influence,ramasesha1991optical}. The intra-phenyl nearest-neighbor hopping is taken to be the same as in BDT, while the inter-phenyl coupling is modeled as $t_\perp(\theta) = (2.49~{\rm eV}) \cos\theta$, where $\theta$ is the dihedral angle between adjacent rings.  
Using the gas-phase torsion angle $\theta_{\rm gas} = 44.4^\circ$ gives $t_\perp(\theta_{\rm gas}) \approx 1.78~{\rm eV}$ \cite{burkle2012conduction,johansson2008torsional}.
Although we employ experimentally motivated parameters, our conclusions do not depend sensitively on their precise~values.

\section{Inclusion of~Dephasing}

Quantum interference effects in molecular conductors can persist even under ambient conditions~\cite{aradhya2012dissecting,arroyo2013signatures,guedon2012observation,markussen2014temperature,liu2018quantum,yang2018quantum,bergfield2024identifying, tang2021reversible,greenwald2021highly}, 
but remain intrinsically sensitive to decoherence from coupling to vibrational, photonic, or~other environmental degrees of freedom.  
To model such processes we employ the B\"uttiker probe approach~\cite{buttiker1986four,buttiker1988coherent}, 
in which fictitious terminals absorb and re-emit carriers stochastically, randomizing phase while preserving global conservation~laws.  

%


Two probe types are considered. A~voltage probe enforces charge conservation ($I_P^{(0)}=0$) but permits finite heat currents, and~thus does not enforce local thermal equilibrium.  
A voltage–temperature probe enforces both charge and heat balance ($I_P^{(0)}=I_P^{(1)}=0$), thereby representing a thermodynamically consistent environment~\cite{engquist1981definition,bergfield2013probing,meair2014local,Stafford2016local,Shastry2015local}.  
The two models coincide in certain symmetric limits but diverge under finite thermal bias or in systems with appreciable thermoelectric response~\cite{eregonarXiv}.  
\JPB{Physically, the~probes act as local measurements on molecular orbitals, effectively introducing incoherent mixing at the attachment site.  
This mechanism directly reduces supernode order without invoking additional degrees~of~freedom.}




As emphasized previously~\cite{bergfield2013probing,bergfield2014thermoelectric,eregonarXiv}, the choice of probe model carries thermodynamic significance.  
A VP fixes the probe temperature externally and enforces only charge conservation, so at finite temperature or in systems with appreciable thermoelectric response it provides, at~best, an~incomplete description of dephasing.  
By contrast, the~VTP enforces full local equilibrium and remains physically consistent across operating conditions.  
In this work we therefore use the VTP throughout our calculations.  
Nevertheless, we retain the VP as an analytic surrogate: its simplicity affords closed-form results for effective Onsager functions in three-terminal circuits, and it reproduces the correct low-energy exponents of the Onsager moments, providing a transparent view of how probe coupling modifies node~order.

\section{Thermoelectric~Observables}
\label{sec:observables}

For a two-terminal circuit with leads $L$ and $R$, the~electrical conductance, thermopower, and~electronic thermal conductance are conveniently expressed in terms of the Onsager functions as
\begin{align}
	\label{eq:G}
	G_{\alpha\beta} &= \left. -e\frac{dI^{(0)}_\alpha}{dV}\right|_{\Delta T = 0} = e^2{\cal L}^{(0)}_{\alpha\beta}, \\[4pt]
	\label{eq:S}
	S_{\alpha\beta} &= - \left.\frac{\Delta V}{\Delta T}\right|_{I^{(0)}_{\alpha}=0} = -\frac{1}{eT_0} \frac{{\cal L}^{(1)}_{\alpha\beta}}{{\cal L}^{(0)}_{\alpha\beta}}, \\[4pt]
	\label{eq:kappa}
	\kappa_{\alpha\beta} &= \left.\frac{dI^{(1)}_\alpha}{d(\Delta T)}\right|_{I^{(0)}_\alpha=0} 
	= \frac{1}{T_0}\left({\cal L}^{(2)}_{\alpha\beta} - \frac{\left[{\cal L}^{(1)}_{\alpha\beta}\right]^2}{{\cal L}^{(0)}_{\alpha\beta}} \right),
\end{align}
where $\alpha,\beta \in \{L,R\}$.  
Here $G$ is the electrical conductance, $S$ is the thermopower (Seebeck coefficient), and~$\kappa$ is the electronic thermal conductance in open-circuit~conditions.

In circuits with more than two electrodes we define the thermal conductance by considering a pure thermal circuit: a temperature bias $\Delta T = T_L - T_R > 0$ is applied while imposing open-circuit conditions on the charge currents, $I_L^{(0)} = I_R^{(0)} = 0$.  
Hence the operational thermal conductance is
\begin{equation}
	\kappa  = \frac{-I_L^{(1)}-I_P^{(1)}}{\Delta T}.
\end{equation}

For a VP, $I_P^{(1)}\neq 0$ and $\kappa$ depends on the probe heat shunt; for a VTP, $I_P^{(1)}=0$ and $\kappa=-I_{L}^{(1)}/\Delta T$, which we report~throughout.

\subsection*{{Device}
~Performance}  

A widely used benchmark for thermoelectric performance is the dimensionless figure of merit~\cite{bell2008cooling,disalvo1999thermoelectric}
\begin{equation}
	ZT_{\rm el} = \frac{S^2 G T_0}{\kappa},
\end{equation}
where $\kappa$ is only the electronic contribution to the thermal conductance.  
In realistic junctions, phonons and other channels provide parallel heat paths that reduce the effective $ZT$\cite{bergfield2010giant}.
For small organic molecules, however, phonon transmission across a metal--molecule interface is typically suppressed due to limited spectral overlap between lead Debye frequencies and molecular vibrational modes.
Accordingly, our focus here is on the electronic part, with~environmental scattering incorporated through voltage--temperature~probes.  

While $ZT$ is a useful rule of thumb, a~more fundamental thermodynamic measure is the thermodynamic efficiency.  
With a small thermal bias, $\Delta T = T_h - T_c$, applied across the junction, the~charge current in linear response is $I_e = G(\Delta V + S\Delta T)$, where $\Delta V = -I_e R_L$ is fixed by the load resistance $R_L$ (or, equivalently, the load ratio $m=GR_L$).  
At this operating point the power delivered to the load is simply
\begin{equation}
P_{\rm out} = I_e^2 R_L.
\end{equation}

For a VTP, which enforces both charge and heat balance, the~hot-side heat input retains the two-terminal form but with probe-renormalized coefficients,
\begin{equation}
	J_{\rm in} = K\,\Delta T + (S T_h)\,I_e - \tfrac{1}{2}(I_e^2/G),
\end{equation}
where $K$ is the total electronic thermal conductance consistent with $ZT = S^2 G T_0/K$.  
The efficiency at finite load then follows as
\begin{equation}
	\eta = \frac{P_{\rm out}}{J_{\rm in}}.
\end{equation}

When reporting efficiencies we normalize to the Carnot value $\eta_C = \Delta  T/T_h$ for the same bias, so that $\eta/\eta_C$ isolates the influence of dephasing and connectivity from temperature scaling.  
All results below correspond to \emph{operating-point} efficiencies at fixed finite $m$; no optimization over load is~performed.


\section{Effective Node-Order Reduction by~Dephasing}
\label{sec:order-reduction-by-dephasing}

\begin{figure}[tb]
	\centering
	\begin{overpic}[width=\linewidth]{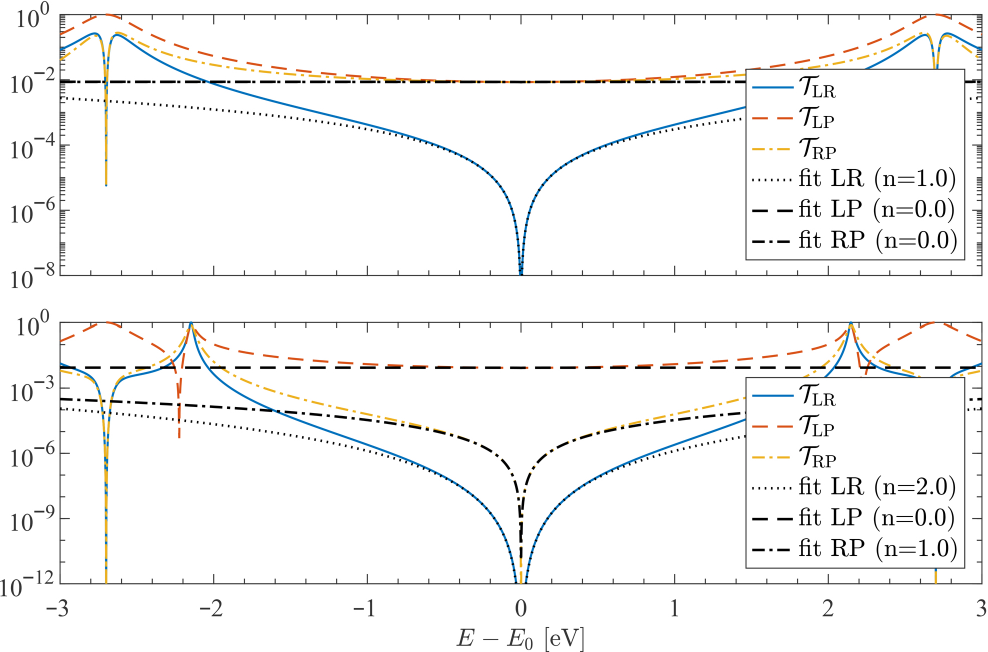}
		\put(12,40){\includegraphics[width=1.1in]{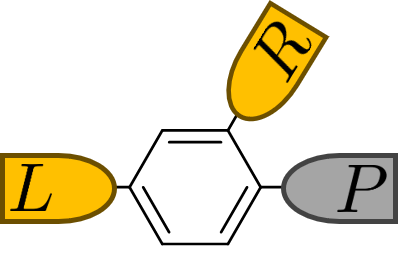}}
		\put(12,8.5){\includegraphics[width=1.2in]{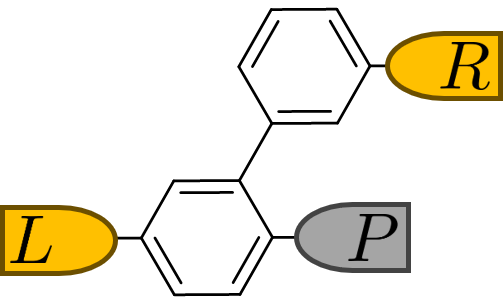}}
	\end{overpic}
	\caption{\textbf{{Transmission functions with a single~probe}
.}
		{The} 
 $\pi$-system {transmissions} 
 for benzene-dithiol (BDT, top) and biphenyl-dithiol (BPDT, bottom) junctions are shown.  
		Solid, dashed, and~dash-dotted curves correspond to the $LR$, $LP$, and~$RP$ channels, respectively.  
		Fits of the form ${\cal T}(E)\propto|E{-}E_{0}|^{2n}$ are overlaid, with~the extracted exponents indicated in the legend.  
		In the BDT junction, the~nodal scaling is quadratic ($n=1$).  
		In the BPDT junction, higher-order behavior emerges: quartic scaling ($n=2$), characteristic of a transmission \emph{{supernode}}, 
        together with nearly flat background contributions from probe-mediated incoherent transport.  
		Because the thermopower is proportional to the energy derivative of the transmission, higher-order nodes (larger $n$) produce correspondingly sharper features and enhanced thermopower responses.  
		Insets show the molecular junction geometries and probe coupling sites.  
		Exponents were extracted over the range $|E{-}E_0|\leq0.3\,$eV.  
		Calculations used the parameters discussed in Section~\ref{sec:hamiltonian} and assume room temperature $T_0=300\,\mathrm{K}$.
	}
	
	\label{fig:fig1_transmission}
\end{figure}

We first investigate the $\pi$-system transport in the vicinity of an interference node (or supernode) for two archetypal junctions: meta-configured Au-1,3-benzenedithiol-Au (BDT) and Au-3,3'-biphenyldithiol-Au (BPDT), each with a single locally coupled probe (shown schematically in the insets of Figure~\ref{fig:fig1_transmission}).  
The NEGF + H\"uckel transmission spectra between all three electrodes are shown in the top and bottom panels of Figure~\ref{fig:fig1_transmission} for BDT and BPDT, respectively.  
For visual comparison, all spectra are shifted so that the node energy is $E_0=0$.  
As expected~\cite{bergfield2009many,bergfield2009thermoelectric,bergfield2010giant}, the coherent left--right channel $\mathcal T_{LR}(E)$ displays a quadratic node in BDT ($n=1$) and a quartic supernode in BPDT ($n=2$).  
The additional probe transmissions, $\mathcal T_{LP}(E)$ and $\mathcal T_{RP}(E)$, depend on connectivity: in BDT the probe is para to the left electrode and ortho to the right, so neither path exhibits a node and both spectra are smooth (i.e., $n=0$).  
In BPDT, by~contrast, the~$L$-$P$ path is para-configured and flat ($n=0$), while the $R$-$P$ path exhibits a quadratic ($n=1$) node, as~indicated by the black fits in Figure~\ref{fig:fig1_transmission}.  

Transport coefficients inherit their scaling from the order of the node. 
The effective exponent $n_{\rm eff}$ therefore provides a direct diagnostic of how dephasing reshapes quantum interference. 
Close to the nodal energy $E_0$, which is detuned from any molecular resonances, each two-terminal transmission channel admits the expansion~\cite{bennett2024quantum}
\begin{equation}
	\mathcal T_{\alpha\beta}(E)
	= A_{\alpha\beta}\,(E-E_0)^{2n_{\alpha\beta}}
	+ \mathcal O\!\big((E-E_0)^{2n_{\alpha\beta}+1}\big),
	\label{eq:T_poly_clean}
\end{equation}
so that the corresponding lowest-order Onsager moments may be expressed as
\begin{equation}
	\mathcal L^{(\nu)}_{\alpha\beta}
	= \frac{A_{\alpha\beta}}{h}\,(k_B T_0)^{2n_{\alpha\beta}+\nu}\,
	F_{\nu,n_{\alpha\beta}}(\delta),
	\qquad
	\delta=\frac{\mu-E_0}{k_B T_0},
	\label{eq:L_scaling_clean}
\end{equation}
with $	F_{\nu,n}(\delta)=\frac{1}{4}\int_{-\infty}^{\infty}
	\epsilon^{2n}\,(\epsilon-\delta)^{\nu}\,
	{\rm sech}^{2}\!\left(\frac{\epsilon-\delta}{2}\right)\,d\epsilon$,
a universal, dimensionless function set solely by the Fermi window. 

Eliminating the probe degrees of freedom from the Onsager matrix yields the Schur complement
\begin{equation}
	\mathbf L^{\rm eff}_{LR}
	= \mathbf L_{LR} - \mathbf L_{LP}\,\mathbf L_{PP}^{-1}\,\mathbf L_{PR},
	\label{eq:schur_vtp}
\end{equation}
where bold symbols denote 2 ${\times}$ 2 Onsager blocks over $\nu=0,1$.
For a VP, $\mathbf L_{PP}$ reduces to the scalar $L^{(0)}_{PP}$, giving the convenient analytic form
\begin{equation}
	\widetilde{\mathcal T}(E)=\mathcal T_{LR}(E)+\frac{\mathcal T_{LP}(E)\,\mathcal T_{PR}(E)}{\mathcal T_{LP}(E)+\mathcal T_{PR}(E)},
	\label{eq:Teff_vp}
\end{equation}
in which the effective transmission is a sum of a coherent $LR$ channel and an incoherent probe-mediated~term. 

For a VTP, a~single energy-local $\widetilde{\mathcal T}(E)$ reproducing both charge and heat currents does not generally exist, since the simultaneous constraints $I_P^{(0)}=I_P^{(1)}=0$ mix energy moments.  
Special cases (e.g.,\ proportional couplings or the narrow-window $k_BT\!\to\!0$ limit) admit such a representation, but~in general one must work directly with Onsager blocks.  
Either way, Equation~\eqref{eq:schur_vtp} shows that $\mathbf L^{\rm eff}_{LR}$ is built from rational combinations of primitive moments, each inheriting a power law $|E-E_0|^{2n_{\alpha\beta}+\nu}$.  
Block inversion cannot increase this power, so the effective near-node exponent is
\begin{equation}
	n_{\rm eff}=\min\!\big(a,\ \max(b,c)\big),
	\label{eq:neff_clean}
\end{equation}
where
\begin{equation}
	{\mathcal T}_{LR} \sim |E-E_0|^{2a}, \qquad 
	{\mathcal T}_{LP}\sim |E{-}E_0|^{2b}, \qquad 
	{\mathcal T}_{PR}\sim |E-E_0|^{2c}.
\end{equation}

A probe, and~by extension any environment faithfully represented by one, can only reduce the sharpness of an interference node; it can never sharpen it.  
Intuitively, phase randomization can destroy interference, but~it cannot create new cancellations.  
Formally, this follows from the Schur-complement structure: dephasing mixes existing channels but does not generate new destructive interference pathways.  
The effective exponent is therefore fixed by the order-selection rule: a single local probe reduces $n_{\rm eff}$ to the order of the strongest bypass, while distributing probes across all orbitals introduces additional incoherent channels that eventually wash out the node entirely.  
Small detunings of the nodal energies (see Appendix~\ref{app:detune_nodes}), alternate probe placements, asymmetric broadenings, or transmission spectra shift numerical prefactors, but~
in all cases the effective exponent $n_{\rm eff}$ is determined by the lowest available order among the coherent and probe-assisted~paths.

The practical importance of $n_{\rm eff}$ lies in its direct control of thermoelectric response.  
As Equation~\eqref{eq:L_scaling_clean} shows, all Onsager blocks inherit the near-node exponent, so transport coefficients such as $G$, $S$, $\kappa$, $ZT$, and~$\eta$ scale parametrically with $n_{\rm eff}$.  
In particular, the~peak thermopower grows nearly linearly with $n_{\rm eff}$, while $ZT_{\rm el}$ and the efficiency $\eta$ are strongly enhanced by higher-order nodes.~\cite{bergfield2010giant}  
Consequently, changes in $n_{\rm eff}$ under dephasing directly translate into the suppression or survival of QI-induced~enhancements.  

\begin{figure}[t]
	\centering
	\begin{minipage}{.49\textwidth}
		\includegraphics[width=\linewidth]{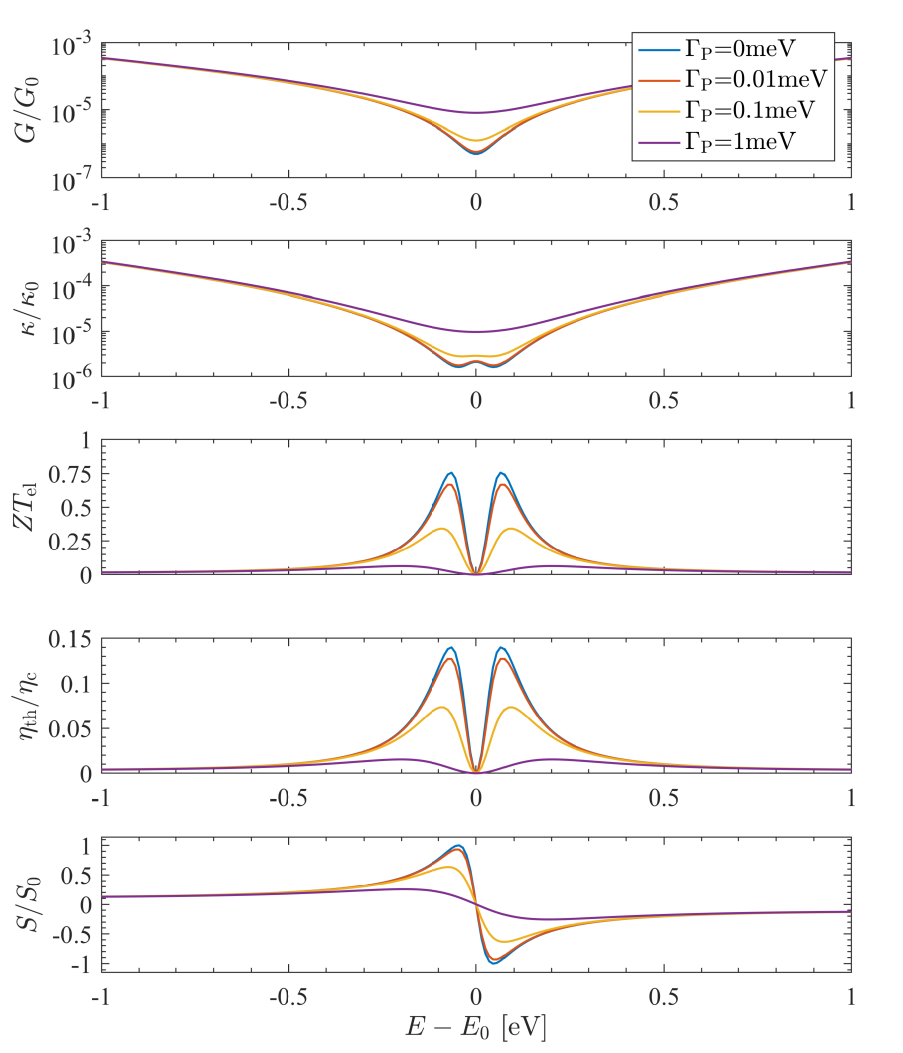}	
		\captionsetup{justification=centering}
		\caption*{({\bf a}) {BDT Junction}
}
	\end{minipage}
	\begin{minipage}{.49\textwidth}
		\includegraphics[width=\linewidth]{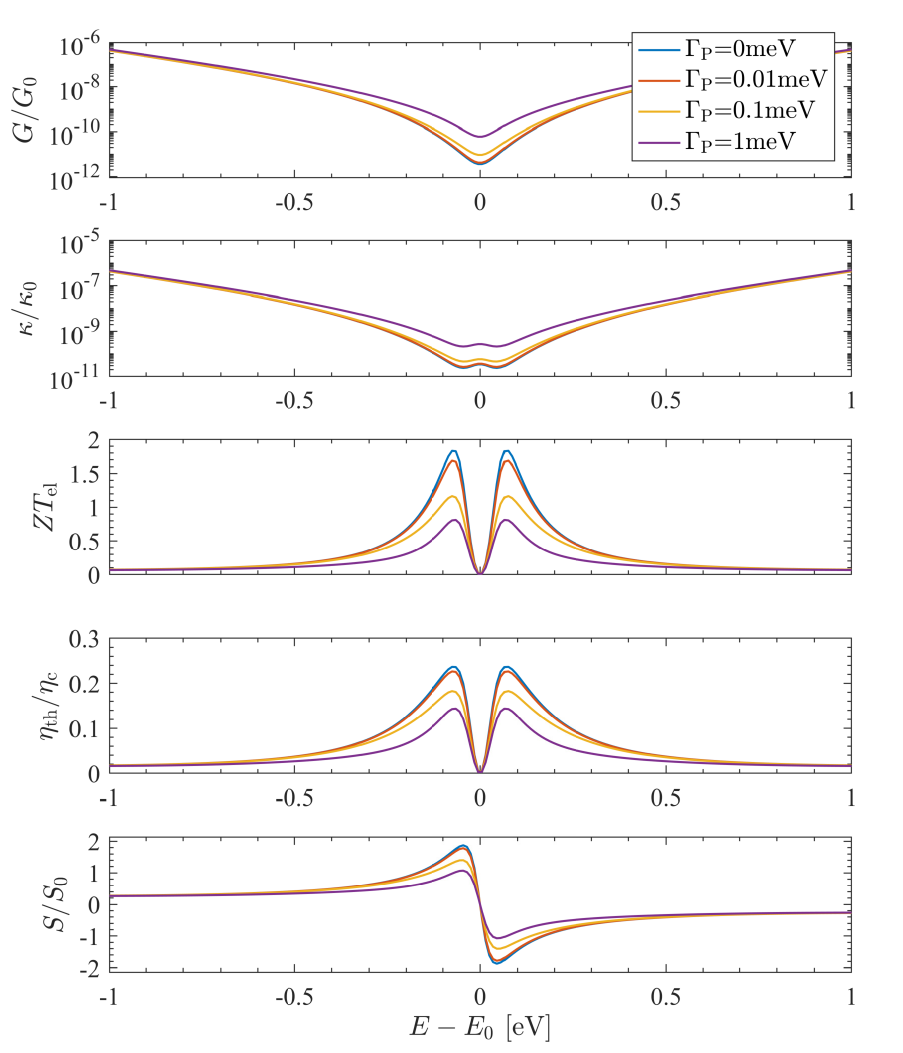}	
		\captionsetup{justification=centering}
		\caption*{({\bf b}) {BPDT Junction}}
	\end{minipage}
	\vspace{3pt}
	\caption{
		\textbf{{Order selection under single-probe VTP dephasing.} 
}  
		{Thermoelectric} 
 {response} 
 of (\textbf{a}) BDT (quadratic node, $n_{\rm coh}=1$) and (\textbf{b}) BPDT (quartic supernode, $n_{\rm coh}=2$) junctions at $T_0=300$\,K.  
		Curves show $G/G_0$, $\kappa/\kappa_0$, $ZT_{\rm el}$, $\eta/\eta_C$, and~$S/S_0$ as functions of energy for several probe couplings $\Gamma_P$.  
		Here $S_0=(k_{B}/e)\pi/\sqrt{3}\approx156\,\mu$V/K, and~$\eta_C$ is the Carnot efficiency defined by the applied $\Delta T$ ($T_{H,C}=T_0 \pm 10$\,K).  
		In BDT, a~probe at the para site generates an incoherent floor ($n_{\rm eff}=0$), while in BPDT the quartic supernode collapses to quadratic order ($n_{\rm eff}=1$).  
		The loss of higher-order scaling suppresses the interference-enhanced thermopower and figure of merit.  
		These trends exemplify the order-selection principle: quadratic nodes persist by saturating to a floor, whereas supernodes collapse to quadratic response under dephasing.  
	}
	\label{fig:fig2_vtp_singleprobe}
\end{figure}

Figure~\ref{fig:fig2_vtp_singleprobe} illustrates these principles for BDT and BPDT junctions with a single local VTP.  
In BDT, the~coherent transmission has order $a=1$; probe-assisted channels contribute $b=c=0$, so $n_{\rm eff}=0$ and a constant background, or~\emph{{floor}}, 
appears at the node.  
In BPDT, the~coherent order $a=2$ collapses to $n_{\rm eff}=1$ without producing a floor, yielding a rapid but continuous crossover in the thermopower and figure of merit.  
In both cases, increasing $\Gamma_P$ suppresses coherence and diminishes interference-induced enhancements of $S$ and $ZT_{\rm el}$.  
Supernodes appear more fragile because their coherent signal is 
smaller near $E_0$, so the incoherent bypass overtakes them at weaker~coupling.  

\vspace{-6pt}

\section{Probe Connectivity: Single vs.\ \boldmath{$N$}-Probe Effects at Fixed Total~Coupling}

Building on the selection rule above, we now examine how different probe connectivities govern the onset of incoherent floor behavior.  
Dephasing probes provide a convenient coarse-grained representation of environmental degrees of freedom.  
A molecular junction embedded in a fluctuating medium, for~example with electrochemical noise, local vibrational baths, or~solvent fluctuations, may exchange particles and heat with many modes that do not couple uniquely to a single orbital.  
To clarify the consequences, we compare two limiting connectivity scenarios.  
In the single-probe case, one VTP couples locally to a designated orbital with strength $\Gamma_{\rm P}$.  
In the $N$-probe case, each orbital couples to its own independent VTP of strength $\Gamma_{\rm P}/N$, so that $\sum_i \Gamma_{P_i}=\Gamma_{\rm P}$ and every orbital relaxes to a separate local voltage and~temperature.

As discussed, a~locally coupled 
probe alters the near-node transmission order according to the selection rule of Equation~\eqref{eq:neff_clean}.  
By contrast, when probes are distributed over all sites, the Schur complement necessarily generates an energy-independent contribution, producing an incoherent floor of the form
\begin{equation}
	\widetilde{\mathcal T}(E)\;\simeq\; B(\Gamma_{\rm P}) \;+\; A\,|E{-}E_0|^{2a} + \cdots,
	\label{eq:T_floor_allsite_restate}
\end{equation}
with $B(\Gamma_{\rm P}) \propto \overline{\mathcal C}(E_0)\,\Gamma_{\rm P}$ in the weak-coupling limit.  
Here $\overline{\mathcal C}(E_0)$ is a geometry-dependent prefactor reflecting how efficiently the probed orbitals overlap both contacts \linebreak  (see Appendix~\ref{sec:appendix_B_linear_in_GammaP}).  

Once a probe-induced floor is present, the~low-energy response is governed entirely by the constant background rather than the coherent order.  
Generally, the~thermopower follows the Mott relation, being proportional to the logarithmic derivative of the transmission near the Fermi level.
Here this derivative scales with the thermal window, while the average transmission is set by $B$, yielding $\max(S) \propto B^{-1/2}$.  
Because both $G$ and $\kappa$ scale with the same $B$, their ratio cancels in $ZT$, leaving $\max(ZT_{\rm el})\propto B^{-1}$.  
Since $B\propto\Gamma_{\rm P}$ in the weak-coupling limit, this produces the universal scalings
%
\begin{equation}
	\max(S) \propto \Gamma_{\rm P}^{-1/2}, \qquad
	\max(ZT_{\rm el})\propto \Gamma_{\rm P}^{-1}.
\end{equation}

These relations hold regardless of whether the underlying node is quadratic or quartic; geometry only enters through the prefactor $\overline{\mathcal C}(E_0)$.

The physical origin of the floor is intuitive: once every orbital is equilibrated by its own probe, 
probe-mediated $L\!\to P_i\!\to R$ paths exist with finite spectral weight even at $E_0$.  
%
Although the order-selection rule still dictates the asymptotic exponent, these probe-mediated bypasses, together with higher-order asymmetries or detuned near–zeros, generate a finite incoherent background. 
In the $N$-probe configuration this background is unavoidable and dominates once $B(\Gamma_{\rm P})$ exceeds the coherent contribution within the thermal~window.



\subsection*{{Small}
-\(\Gamma_{\rm P}\): A Single Probe Maximally Dephases}

At weak coupling, probes act independently and their contributions to the effective $LR$ block are additive.  
Denoting by $\mathbf L_{\alpha\beta}$ the $2\times2$ Onsager blocks over $\nu=0,1$, one finds for a probe on site $i$,
\begin{equation}
	\mathbf L_{LP_i}=\Gamma_{P_i}\,\mathbf A_i+\mathcal O(\Gamma_{P_i}^2),\quad
	\mathbf L_{P_iP_i}=\Gamma_{P_i}\,\mathbf C_i+\mathcal O(\Gamma_{P_i}^2),\quad
	\mathbf L_{P_iR}=\Gamma_{P_i}\,\mathbf B_i+\mathcal O(\Gamma_{P_i}^2),
\end{equation}
with $\mathbf C_i$ positive definite and $\mathbf A_i,\mathbf B_i,\mathbf C_i$ independent of $\Gamma_{P_i}$ at leading order.  
Inserting into Equation~\eqref{eq:schur_vtp} yields the first correction to the effective $LR$ block,
\begin{equation}
	\Delta \mathbf L_{LR}^{\rm eff}
	=\sum_i \Gamma_{P_i}\,\mathbf K_i+\mathcal O(\Gamma_{\rm P}^2),
	\qquad
	\mathbf K_i\equiv \mathbf A_i\,\mathbf C_i^{-1}\,\mathbf B_i .
	\label{eq:probe_linear}
\end{equation}
The consequence for thermopower can be written as the initial slope
\begin{equation}
	\frac{dS}{d\Gamma_{\rm P}}\Big|_{\Gamma_P=0}
	=-\frac{1}{eT}\;
	\frac{\sum_i w_i\,K_{i}^{(1)}\,\mathcal L^{(0)}_{LR}
		- \mathcal L^{(1)}_{LR}\,\sum_i w_i\,K_{i}^{(0)}}
	{\big(\mathcal L^{(0)}_{LR}\big)^{2}},
	\label{eq:S_linear_slope}
\end{equation}
with weights $w_i=\Gamma_{P_i}/\Gamma_{\rm P}$ (\(\sum_i w_i=1\)).  
%
Because this expression is affine in the $\{w_i\}$, the~steepest suppression of $|S|$ at fixed $\Gamma_{\rm P}$ occurs when all coupling is placed on a single orbital.  
The probe opens a direct incoherent pathway through that orbital, whose strength is governed by molecular symmetry and the local spectral weight at site $i$.  
Concentrating the coupling on the orbital that maximizes the probe-mediated transmission therefore produces the strongest dephasing effect, directly reducing the effective node order.  
By contrast, probes attached to symmetry-dark orbitals, i.e.,~those which carry vanishing spectral weight at $E_0$, contribute negligibly and are far less disruptive~\cite{cardamone2006controlling}.

\begin{figure}[t]
	\centering
	\begin{minipage}{.49\textwidth}
		\includegraphics[width=\linewidth]{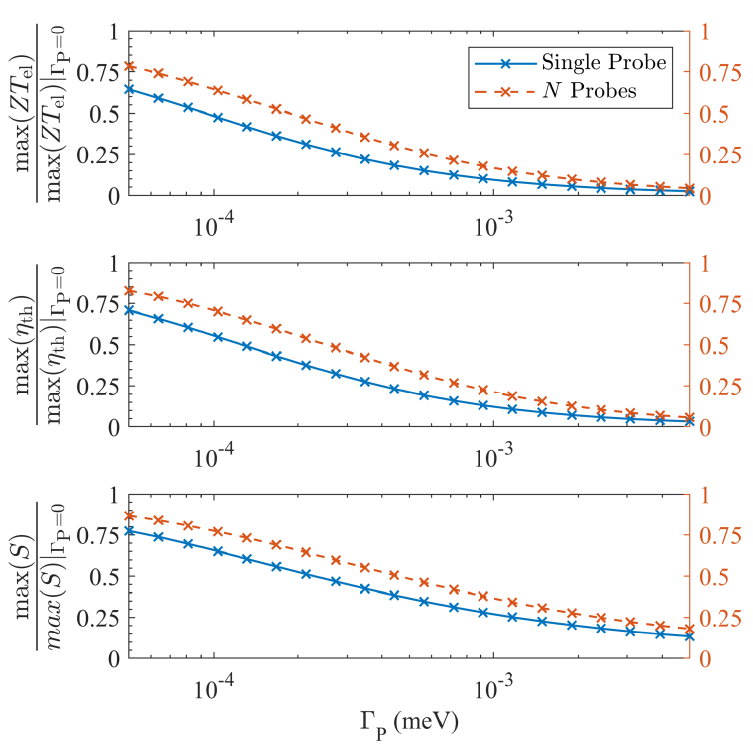}	
		\captionsetup{justification=centering}
		\caption*{({\bf a}) BDT Junction}
	\end{minipage}
	\begin{minipage}{.49\textwidth}
		\includegraphics[width=\linewidth]{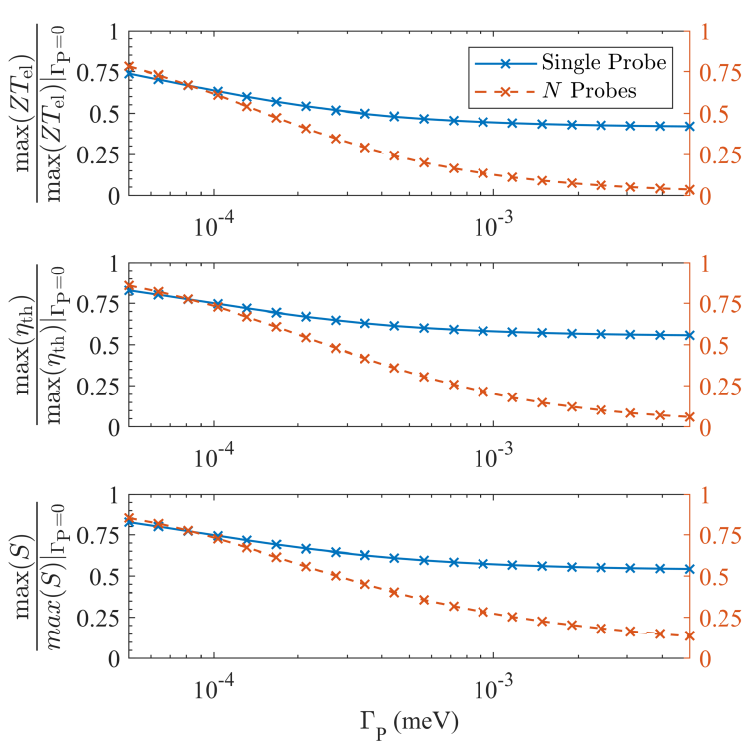}	
		\captionsetup{justification=centering}
		\caption*{({\bf b}) BPDT Junction} 
	\end{minipage}
		\vspace{3pt}
	\caption{\textbf{{Reduction of node enhancements versus total probe~coupling.}
}
		{Maxima} 
 (over~$E$) {of} 
 $ZT_{\rm el}$, $\eta_{\rm th}$, and~$S$ at $T=300$\,K, normalized to their coherent peaks, are plotted against the \emph{total} probe strength $\Gamma_{\rm P}$. Blue curves show a single local probe of strength $\Gamma_{\rm P}$; orange curves show $N$ independent local probes (one per site), each of strength $\Gamma_{\rm P}/N$. At~small $\Gamma_{\rm P}$ the single-probe traces fall at least as rapidly, consistent with order selection in which a supernode collapses to quadratic order without creating a floor. At~larger $\Gamma_{\rm P}$ the all-site traces cross below, reflecting the emergence of a genuine incoherent floor and the ensuing $S_{\max}\!\propto\!\Gamma_{\rm P}^{-1/2}$, $ZT_{\max}\!\propto\!\Gamma_{\rm P}^{-1}$ scaling. The~crossover scale follows the same temperature trend as the order-selection scale $\Gamma_c$ but is observable- and geometry-dependent.}
	\label{fig:ZT_eta_S_vs_GammaP_for_SingleProbe_and_NProbes}	
\end{figure}

This behavior is evident in Figure~\ref{fig:ZT_eta_S_vs_GammaP_for_SingleProbe_and_NProbes}.  
For the BDT node, shown in panel (a), a~single local probe depresses the \emph{{normalized}} 
thermopower and efficiency more strongly, across nearly the entire $\Gamma_{\rm P}$ range, than~$N$ smaller probes of the same total strength.  
For BPDT, shown in panel (b), a~single probe immediately collapses the supernode ($2\!\to\!1$) without creating a floor, again producing a steeper normalized suppression.  
Absolute peak values can nevertheless remain large, depending on prefactors, but~the trend is clear: concentrating $\Gamma_{\rm P}$ on a single orbital with significant transport maximizes the dephasing~effect.

At larger $\Gamma_{\rm P}$, however, the~balance shifts.  
When probes are distributed over all orbitals, the~cumulative floor $B(\Gamma_{\rm P})$ grows linearly with the number of available bypasses.  
Once this floor overtakes the dispersive contribution, many-probe geometries suppress $S$, $ZT$, and~$\eta$ more efficiently than a single probe, leading to the crossings seen in Figure~\ref{fig:ZT_eta_S_vs_GammaP_for_SingleProbe_and_NProbes}.  
In other words, single-site coupling dominates in the weak-dephasing regime, whereas $N$-probe coupling dominates once the incoherent background becomes~appreciable.


The resulting contrast is summarized in Figure~\ref{fig:-}. With one probe per site, each coupled at strength $\Gamma_{\rm P}/N$ so that the total coupling is fixed, the~\emph{{normalized}} 
suppression of $S$, $ZT_{\rm el}$, and~$\eta$ becomes nearly identical for BDT (quadratic node) and BPDT (quartic supernode).  
This reflects the fact that, once a probe-induced floor is present, the~\emph{{fractional}} 
reduction of the thermoelectric response is governed primarily by $B(\Gamma_{\rm P})$ rather than by the underlying coherent order.  
Absolute values can still differ substantially---supernodes retain their larger coherent-limit enhancements until the floor dominates---but~the \emph{{shape}} 
of the decay becomes order-independent eventually.  
The modest residual curvature differences between the two molecules reflect only geometry-dependent prefactors in $\overline{\mathcal C}(E_0)$, not a distinct order-selection mechanism.  
Thus, while single-site probes reveal the fragility of supernodes through immediate order reduction, all-site dephasing renders the \emph{{fractional suppression}} 
effectively order-agnostic.

\vspace{-6pt}
\begin{figure}[t]
	\centering 
	\begin{minipage}{.6\textwidth}
		\includegraphics[width=\linewidth]{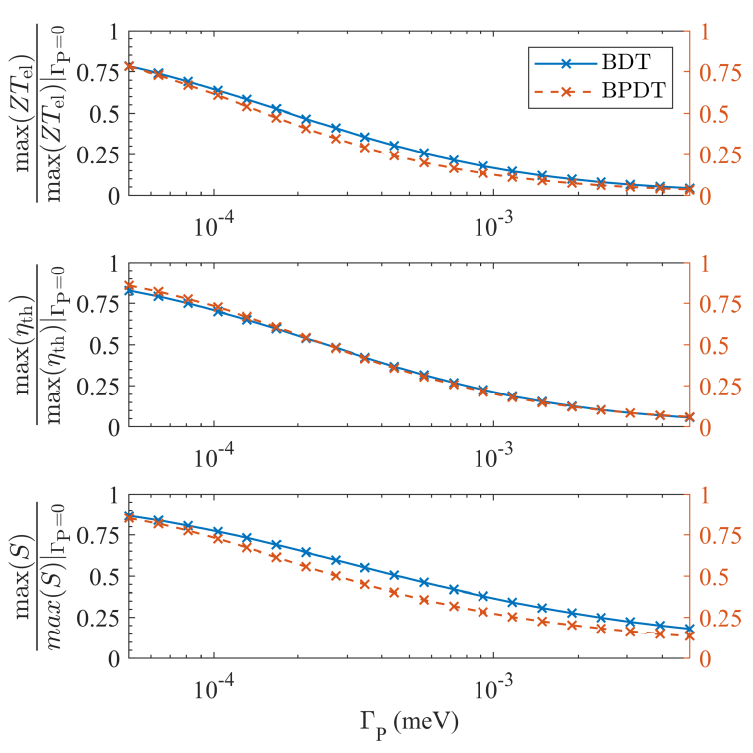}	
	\end{minipage}
	\caption{\textbf{{All-site dephasing renders fractional suppression~order-agnostic.} 
}  
		{With} 
 {one} 
 probe per site (each coupled with $\Gamma_{\rm P}/N$, so the total is $\Gamma_{\rm P}$), the~\emph{{normalized}} 
 decays of $ZT_{\rm el}$, $\eta$, and~$S$ versus $\Gamma_{\rm P}$ are nearly indistinguishable for BDT (quadratic node; left vertical axis) and BPDT (quartic supernode; right vertical axis) at $T=300$\,K.  
		This demonstrates that once a probe-induced floor is present, the~\emph{{fractional}} 
        reduction of the response is governed primarily by $B(\Gamma_{\rm P})$ and is largely insensitive to the coherent order.  
		Absolute values can still remain larger for supernodes, reflecting their higher coherent-limit enhancements, but~the \emph{{shape}} 
        of the decay is universal.  
		The modest residual curvature differences reflect geometry-dependent prefactors 
		rather than a distinct order-selection mechanism. }
	\label{fig:-}
	
	\label{fig:-}	
\end{figure}

%


\section{Order-Dependent Sensitivity to~Dephasing}

A central question in this work is whether higher-order interference supernodes are intrinsically more fragile to dephasing than ordinary quadratic nodes.  
Because both the conductance and thermal conductance inherit the near-node exponent, their temperature scaling provides a direct window into how coherence is degraded.  
Formally, the~Onsager relation of Equation~\eqref{eq:L_scaling_clean} implies
\begin{equation}
	G \propto (k_B T)^{2n_{\rm eff}}, \qquad
	\kappa_e \propto (k_B T)^{2n_{\rm eff}+1},
	\label{eq:node_scalings}
\end{equation}
so the log--log slopes of $G$ and $\kappa_e$ at $E_0$ give a direct measure of the effective node order $n_{\rm eff}$.
We therefore investigate supernode fragility by tracking how the effective nodal order, extracted from the slopes of $G$ and $\kappa_e$, evolves as a function of probe coupling~strength.

\begin{figure}[t]
	\begin{minipage}{.49\textwidth}
		\includegraphics[width=\linewidth]{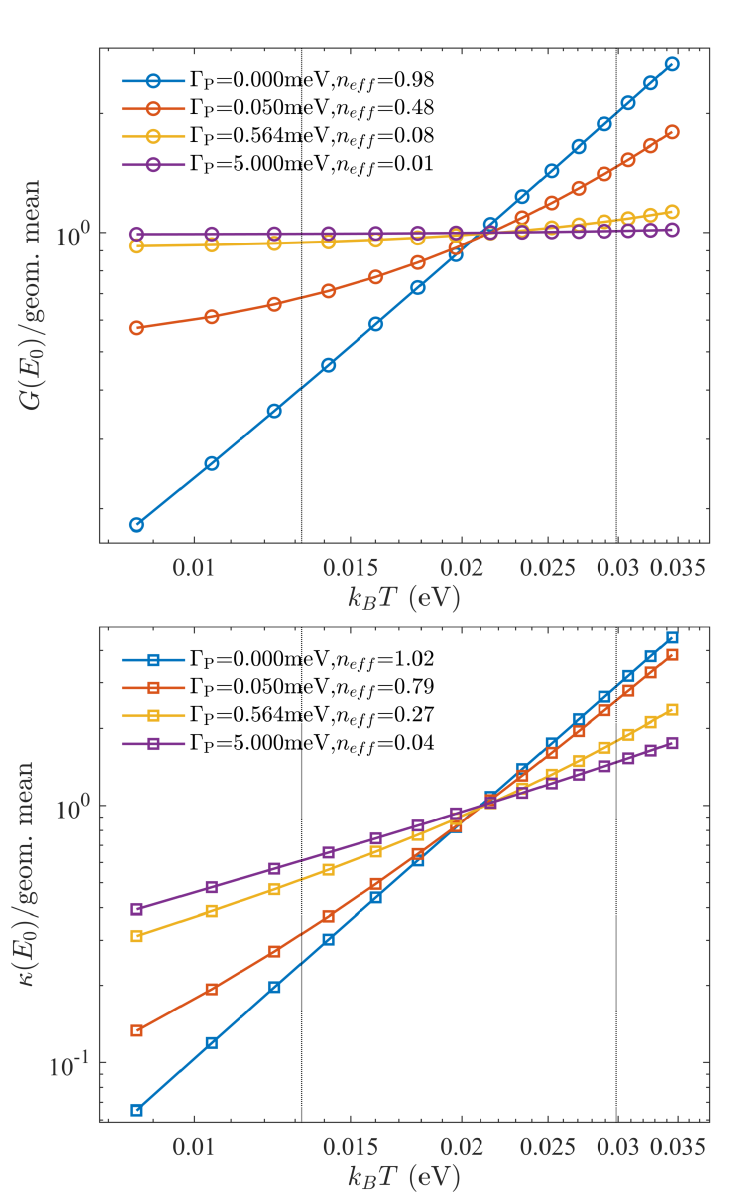}
		\captionsetup{justification=centering}	
		\caption*{({\bf a}) BDT Junction}
	\end{minipage}
	\begin{minipage}{.49\textwidth}
		\includegraphics[width=\linewidth]{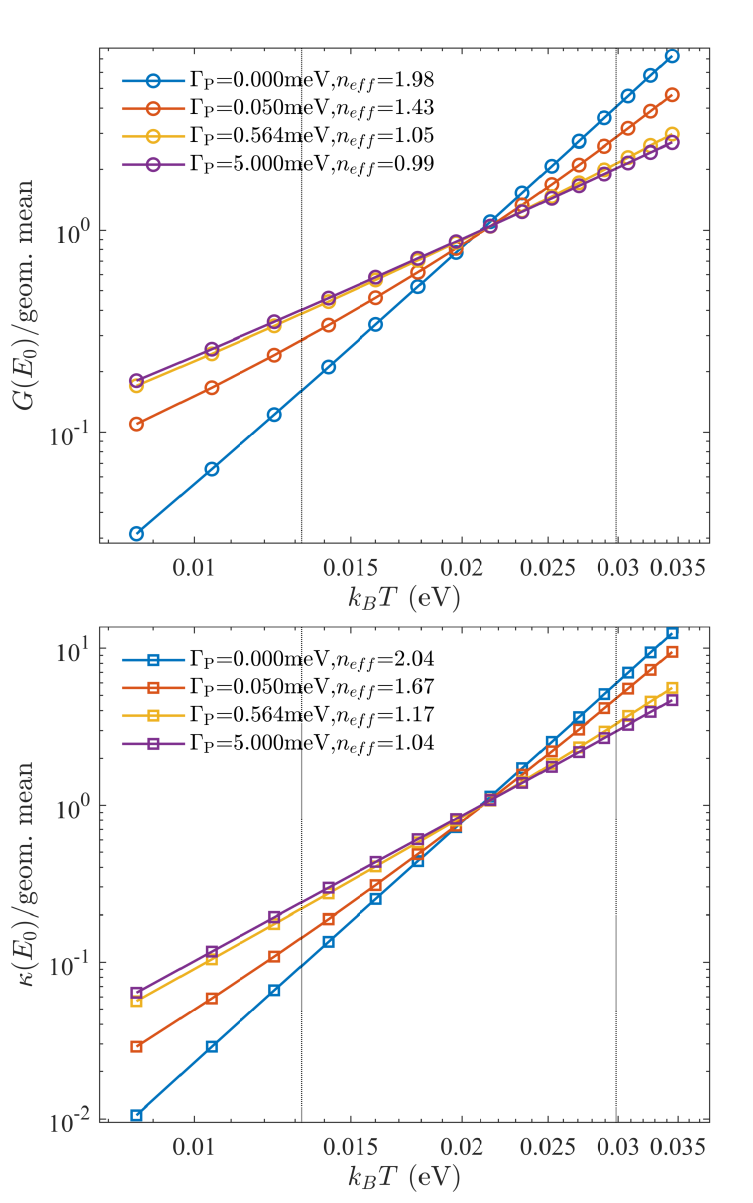}	
		\captionsetup{justification=centering}
		\caption*{({\bf b}) BPDT Junction} 
	\end{minipage}
	\vspace{3pt}
	\caption{\textbf{{Node scaling identifies the effective order under~dephasing.} 
}
		{Log--log} 
 {plots} 
 of conductance $G(E_0)$ and electronic thermal conductance $\kappa(E_0)$ versus $k_B T$ for several single probe coupling strengths $\Gamma_{P}$.  
		Each trace is normalized by its geometric mean over the fit window (dotted vertical lines), so that vertical offsets are removed and slopes directly yield the effective order $n_{\rm eff}$.  
		Values extracted from $G$ and $\kappa_e$ agree within uncertainty.  
		In panel (\textbf{a}), the~BDT (benzene) junction has $(a,b,c)=(1,0,0)$, giving $n_{\rm eff}=1$ in the coherent limit. Any $\Gamma_{P}>0$ introduces an energy-independent floor, driving $n_{\rm eff}\to 0$ with $G\!\sim T^0$ and $\kappa_e\!\sim T^1$.  
		In panel (\textbf{b}), the~BPDT (biphenyl) junction has $(a,b,c)=(2,0,1)$, yielding $n_{\rm eff}=2$ ($G\!\sim T^4$, $\kappa\!\sim T^5$) in the coherent limit. As~$\Gamma_{P}$ is increased this quartic supernode collapses to quadratic order, $n_{\rm eff}=1$, with~$G\!\sim T^2$ and $\kappa_e\!\sim T^3$.  
		In both cases the probe coupling is weak compared to $\Gamma_{L,R}$, so the coherent spectrum is otherwise unaltered; the probe acts only to select the effective order. \JPB{The single probe junction configurations are indicated in  Figure~\ref{fig:fig1_transmission}}. }
	
	\label{fig:node_scaling}
\end{figure}

The conductance $G(E_0)$ and electronic thermal conductance $\kappa(E_0)$ are shown in \linebreak  Figure~\ref{fig:node_scaling} as functions of $k_B T$ for several single probe couplings $\Gamma_{\rm P}$.  
Each trace is normalized by its geometric mean over the fit window (dotted vertical lines), so that offsets are removed and the slopes directly reveal the effective order $n_{\rm eff}$.  
Values extracted independently from $G$ and $\kappa_e$ agree within uncertainty, confirming the robustness of this~diagnostic.

Panel (a) illustrates the BDT junction $(a,b,c)=(1,0,0)$.  
In the coherent limit the slope corresponds to $n_{\rm eff}=1$, as~expected for a quadratic node.  
Any finite $\Gamma_{\rm P}$, however, introduces an energy-independent bypass that drives $n_{\rm eff}\!\to 0$, yielding $G\!\sim T^0$ and $\kappa_e\!\sim T^1$.  
Panel (b) shows the BPDT supernode $(a,b,c)=(2,0,1)$, which in the coherent limit yields $n_{\rm eff}=2$.  
Here even an infinitesimal $\Gamma_{\rm P}$ collapses the quartic scaling to quadratic ($n_{\rm eff}=1$) without generating a floor, reflecting the immediate fragility of the supernode.
An analogous analysis applies to $N$-probe geometries (Appendix~\ref{app:Nprobe_scaling}).  
In this case, distributing probes across all orbitals produces an incoherent floor that enforces $n_{\rm eff}\!\to 0$ as $\Gamma_{\rm P}$ increases, rendering the suppression effectively order-agnostic once the floor~dominates.

\vspace{-6pt}

To quantify these crossovers, we fit the extracted slopes to the interpolation
\begin{equation}
	n_{\rm eff}(\Gamma_{\rm P})
	= n_\infty + \frac{n_0-n_\infty}{\big(1+\Gamma_{\rm P}/\Gamma_c\big)^{\beta}},
	\label{eq:neff_general}
\end{equation}
where $n_0$ and $n_\infty$ denote the coherent and dephased asymptotes, respectively \linebreak  (see Appendix~\ref{app:neff}).  
From the single-probe spectra we obtain
\begin{align}
	\Gamma_c^{\rm BDT} &= 0.0794 \pm 0.0057~\mathrm{meV}, \qquad \beta^{\rm BDT}=0.86 \pm 0.03, \\[2pt]
	\Gamma_c^{\rm BPDT} &= 0.0540 \pm 0.0044~\mathrm{meV}, \qquad \beta^{\rm BPDT}=0.91 \pm 0.03.
\end{align}

The fitted exponents remain close to $\beta\simeq 1$, as~expected for probe-induced incoherent channels that scale linearly with $\Gamma_{\rm P}$.  
The crucial difference lies in the crossover scales $\Gamma_c$.  
Because the coherent transmission of a supernode is parametrically suppressed near $E_0$, even a weak bypass rapidly overwhelms it.  
The biphenyl supernode therefore collapses at a probe strength roughly $1.5\times$ smaller than that required to quench the benzene node.  
This is the precise sense in which supernodes are “more fragile’’: not that their collapse is sharper, but~that it occurs at parametrically smaller $\Gamma_{\rm P}$.  
In other words, their enhanced sensitivity is rooted in spectral weight, not in the nature of the dephasing~itself.

Extending the same analysis to $N$-probe geometries gives
\begin{align}
	\Gamma_c^{\rm BDT}&=0.14 \pm 0.010~\mathrm{meV}, \quad \beta^{\rm BDT}=0.81 \pm 0.03, \\[2pt]
	\Gamma_c^{\rm BPDT}&=0.15 \pm 0.012~\mathrm{meV}, \quad \beta^{\rm BPDT}=1.09 \pm 0.04,
\end{align}
indicating that once probes are distributed across all orbitals, both junctions develop \JPB{a nearly indistinguishable} 
incoherent floor. 
With this floor, 
the \emph{fractional} suppression of $\max(S)$, $\max(ZT_{\rm el})$, and~$\max(\eta)$ is essentially order-agnostic (cf. Figure~\ref{fig:-}), although~geometry sets the prefactors.

We can now return to the central question of this work.  
Supernodes are indeed more sensitive to \emph{local} dephasing. However, once the environment acts collectively, as~modeled by $N$-probe configurations, the~distinction between quadratic nodes and supernodes effectively disappears.  
The lesson is twofold.  
First, although~higher-order supernodes provide enhanced thermoelectric response, that enhancement is eroded by weaker coupling than a quadratic node.
Second, the~way the environment couples, through a single dominant orbital or through many, dictates whether this erosion is abrupt (order reduction) or gradual (floor building).  
Thus, it is not merely the presence of dephasing, but~the geometry of its coupling, that determines whether supernode-based thermoelectric enhancement~survives.

\section{Conclusions}  

We have shown that dephasing always reduces, but~never sharpens, quantum interference nodes.  
A probe either leaves the node order unchanged or collapses a supernode to lower order.  
This asymmetry reflects the Schur-complement structure of the probe formalism: effective Onsager blocks inherit the lowest available power law, so incoherent mixing can only reduce, not enhance, destructive interference.  
In other words, probes encode incoherent pathways but do not generate new interference~routes.  

This behavior is summarized by a simple \emph{{order-selection principle}}: 
the effective near-node order is given by the minimum of the coherent exponent and the largest exponent accessible through probe-mediated transport.  
In practice, a~quartic supernode collapses quickly to quadratic order under even weak local perturbations, while a quadratic node preserves its form until incoherent bypass channels introduce a true floor.  
The associated crossover is characterized by two fitted parameters: $\Gamma_c$, the~probe strength at which incoherent processes overtake the coherent node, and~$\beta$, which controls the sharpness of the collapse.  
We find $\beta\simeq 1$ in both single-probe and $N$-probe connectivities, consistent with probe-induced channels that scale linearly with $\Gamma_{\rm P}$, while $\Gamma_c$ captures the enhanced sensitivity of supernodes under local~coupling.  

The reduction of a node's effective order is continuous: within the thermal window the coherent contribution scales as $(k_BT)^{2n}$ while probe-mediated terms grow with $\Gamma_P^\beta$, so $n_{\rm eff}$ decreases smoothly rather than discontinuously.  
Both quadratic and quartic nodes degrade at comparable rates with increasing $\Gamma_{\rm P}$; the distinction is that supernodes cross over at weaker probe strengths, reflecting their greater fragility in the order-selection sense rather than a faster decay~rate.

Probe connectivity plays an equally important role.  
For fixed total coupling, a~single local probe reduces the order but does not produce a floor, whereas distributing the same strength across multiple sites inevitably builds one.  
Once present, this floor enforces the scaling, 
$\max(S)\propto \Gamma_{\rm P}^{-1/2}$ and $\max(ZT_{\rm el})\propto \Gamma_{\rm P}^{-1}$, independent of the initial coherent order.  
In this regime the degradation of thermopower, efficiency, and~figure of merit becomes order-agnostic, governed primarily by the prefactor of the incoherent floor. 

The stability of supernodes, and~the QI-driven enhancements they support, therefore depends not only on the overall coupling strength but also on how the environment connects to molecular orbitals.  
Probes make this dependence explicit, revealing when supernodes retain their advantage and when interference collapses to universal scaling laws.  
In this sense, dephasing becomes a design principle: robustness can be maximized by engineering environmental couplings or by exploiting molecular symmetries that preserve nodal pathways, suggesting practical strategies for realizing quantum-enhanced \linebreak  {thermoelectric~materials.} 


\funding{{This} 
 research was graciously supported by the National Science Foundation under award number~QIS-2412920.}

%



\dataavailability{{The} 
 original contributions presented in the study are included in the article; further inquiries can be directed to the corresponding author.}

\conflictsofinterest{The author declares no conflicts of~interest.} 

\appendixtitles{yes} 
\appendixstart
\appendix
\section{Order Selection with \boldmath{$N$} Local~Probes}
\label{app:Nprobe_scaling}


To confirm that the node order-selection rule of Equation~\eqref{eq:neff_clean} is not an artifact of single-probe coupling, 
we repeated the log--log scaling analysis with $N$ independent VTPs, one attached to each molecular orbital.  
Each probe enforces local equilibration (\(I_P^{(0)}=I_P^{(1)}=0\)) with coupling $\Gamma_{\rm P}/N$, so that the total strength remains fixed at $\Gamma_{\rm P}$.  
This configuration represents the strongest form of dephasing, since every orbital exchanges particles and heat with its own~bath.  

\begin{figure}[t]
	\begin{minipage}{.49\textwidth}
		\includegraphics[width=\linewidth]{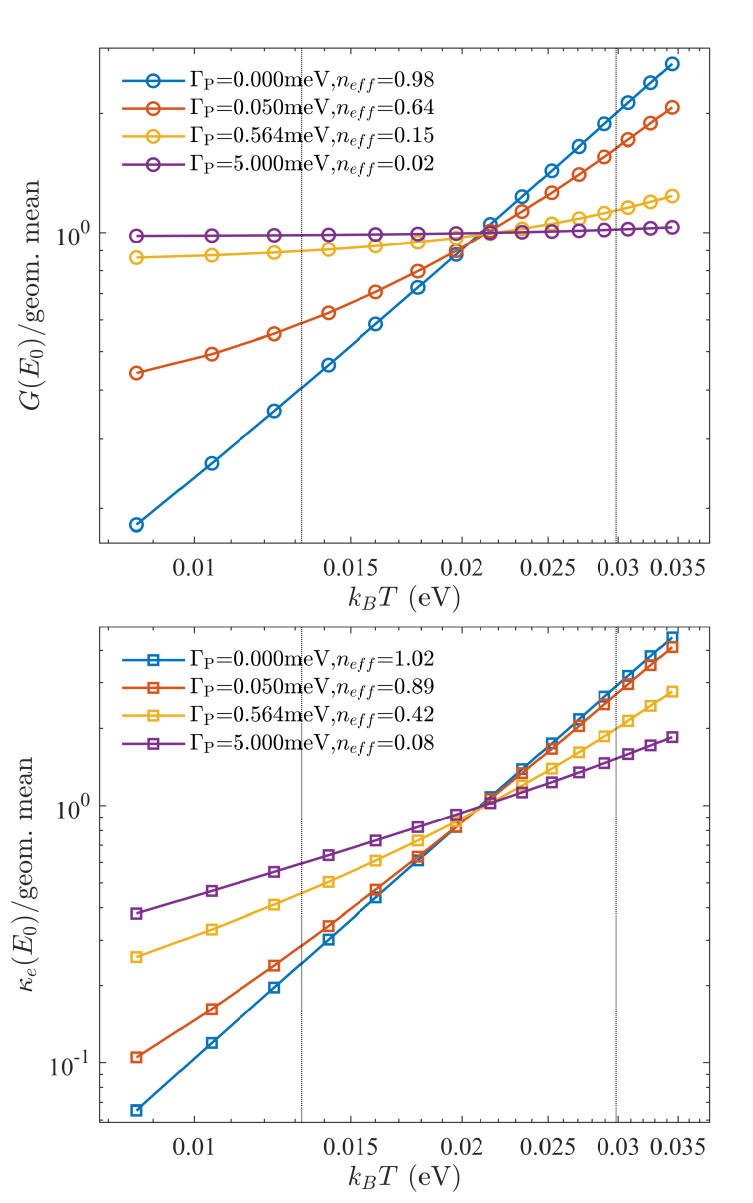}	
		\captionsetup{justification=centering}
		\caption*{({\bf a}) BDT Junction ($N=6$ probes)}
	\end{minipage}
	\begin{minipage}{.49\textwidth}
		\includegraphics[width=\linewidth]{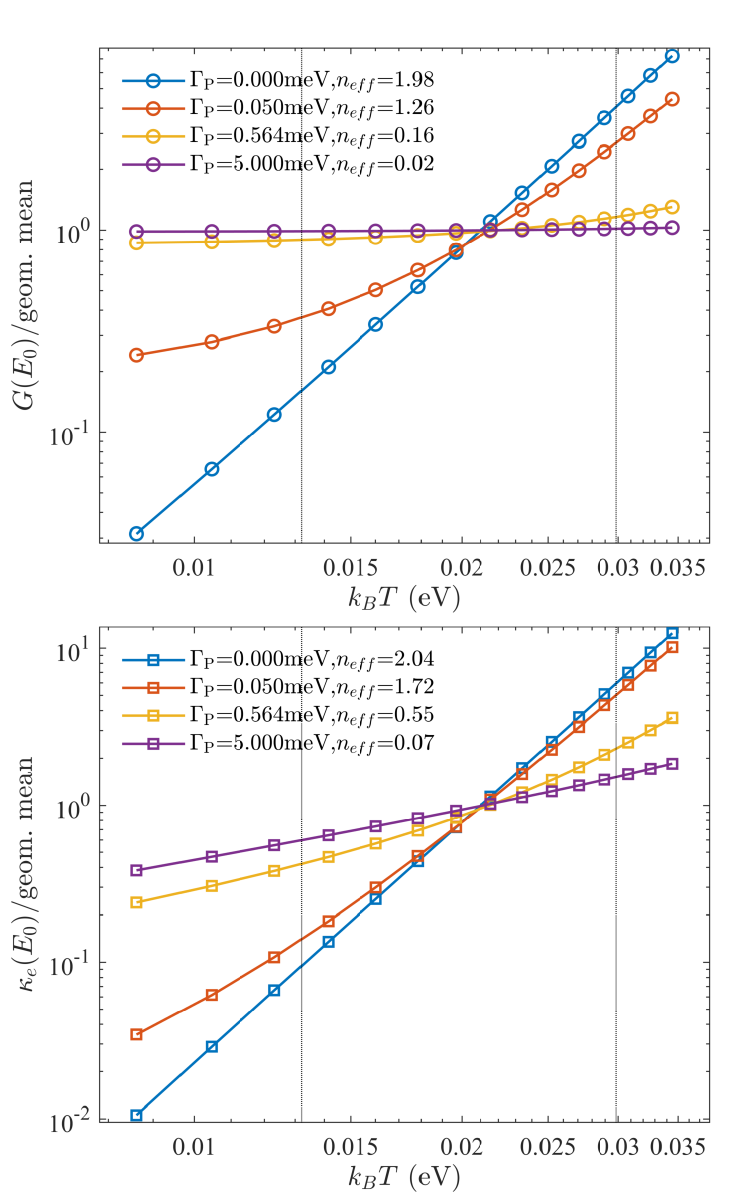}
			\captionsetup{justification=centering}
		\caption*{({\bf b}) BPDT Junction ($N=12$ probes)} 
	\end{minipage}
	\vspace{3pt}
	\caption{\textbf{{Effect of probe connectivity on node~scaling.}
}
		{Same} 
 {analysis} 
 as in Figure~\ref{fig:node_scaling}, but~with $N$ independent VTPs locally coupled and equilibrated to every molecular site.  
		Each probe is allowed to relax locally to its own chemical potential and temperature, thereby providing the strongest form of dephasing.  
		Panel (\textbf{a}) shows the {BDT junction} 
 and panel (\textbf{b}) {the BPDT junction.}  
		The qualitative order-selection behavior is unchanged in this case, however.}
	
	\label{fig:node_scaling_allsites}	
\end{figure}

Figure~\ref{fig:node_scaling_allsites} shows the resulting scaling of $G(\mu_0)$ and $\kappa_e(\mu_0)$ with $k_B T$ for BDT and BPDT.  
In both molecules the effective order $n_{\rm eff}$ is again determined by the lowest available exponent among the coherent and probe-assisted channels, exactly as predicted by the selection rule.  
No evidence is found for sharpening or enhancement of interference nodes under many-site coupling: the incoherent probe pathways always reduce the effective~order.  

The principal difference from the single-probe case is quantitative rather than qualitative.  
Because all orbitals are equilibrated, additional two-step paths ($L\!\to P_i\!\to R$) are available, generating an incoherent background that appears as a finite floor at the node [cf.\ Equation~\eqref{eq:T_floor_allsite_restate}].  
This background dominates more rapidly as $\Gamma_{\rm P}$ increases, but~the near-node scaling exponents extracted from Figure~\ref{fig:node_scaling_allsites} follow the same order-selection principle already discussed in the main text.  
Thus, the $N$-probe geometry confirms that the order-selection rule is general: dephasing can only lower the effective order of a node or introduce a floor, never sharpen~it.  

\section{Distributed Probes and the Linear~Floor}
\label{sec:appendix_B_linear_in_GammaP}
Consider an $N$-probe VTP configuration in the wide-band, weak-coupling limit with local probe couplings
$\Gamma_{P_i}\ll \Gamma_{L},\Gamma_{R}$ at orbitals $\{|i\rangle\}$ and total $\Gamma_{\rm P}=\sum_i \Gamma_{P_i}$.
Let $G_0^{r/a}(E)$ denote the retarded/advanced Green's functions of the \emph{{unprobed}} 
junction. Then at the nodal energy $E_0$,
the effective two-terminal transmission, after eliminating the probes by the Schur complement, acquires an
energy-independent term,
\begin{equation}
	\widetilde{\mathcal T}(E_0)
	\;=\;
	\mathcal T_{LR}^{\rm coh}(E_0)
	\;+\;
	\sum_{i=1}^N \Gamma_{P_i}\, \mathcal C_i(E_0)
	\;+\;
	\mathcal O(\Gamma_{\rm P}^2),
	\label{eq:linear_floor_statement}
\end{equation}
with nonnegative coefficients
\begin{equation}
	\mathcal C_i(E) \;=\;
	\frac{
		\alpha_{iL}(E)\,\alpha_{iR}(E)
	}{
		\alpha_{iL}(E)+\alpha_{iR}(E)
	},
	\qquad
	\alpha_{i\alpha}(E)\equiv \langle i|G_0^{a}(E)\,\Gamma_\alpha\,G_0^{r}(E)|i\rangle \ge 0,
	\label{eq:C_i_def}
\end{equation}
and $\Gamma_\alpha$, the usual lead broadenings. 
 In~particular, for~equal splitting
$\Gamma_{P_i}=\Gamma_{\rm P}/N$,
\begin{equation}
	B(\Gamma_{\rm P}) \;\equiv\; \widetilde{\mathcal T}(E_0)-\mathcal T_{LR}^{\rm coh}(E_0)
	\;=\;
	\Gamma_{\rm P}\,\overline{\mathcal C}(E_0) \;+\; \mathcal O(\Gamma_{\rm P}^2),
	\label{eq:B_linear}
\end{equation}
where
\begin{equation}
	\overline{\mathcal C}(E_0)=\frac{1}{N}\sum_{i=1}^N \mathcal C_i(E_0)
\end{equation}
is the geometry-dependent prefactor that converts the total probe coupling $\Gamma_P$ into an effective incoherent~background.

In linear response, eliminating the probes with the VTP constraints ($I_p^{(0)}=I_p^{(1)}=0$) yields the effective Onsager block~\cite{eregonarXiv}
\begin{equation}
	\widetilde{\mathcal L}^{(\nu)}_{LR}
	\;=\;
	\mathcal L^{(\nu)}_{LR}
	\;+\;
	\mathcal L^{(\xi)}_{L P}\,
	\big[-\mathcal L^{(0)}_{PP}\big]^{-1}
	\mathcal L^{(\nu-\xi)}_{P R},
	\label{eq:Schur_general}
\end{equation}
for any $\xi$ (we choose $\xi=0$ for conductance), where $P$ indexes the probe set.
To leading order in $\Gamma_{P_i}$,
\begin{align}
	\mathcal L^{(0)}_{L\,i} &=
	\frac{1}{h}\!\int\! dE\,
	\Big(-\frac{\partial f_0}{\partial E}\Big)\,
	\mathrm{Tr}\!\big[\Gamma_L G_0^{r}(E)\,\Gamma_{P_i}\,G_0^{a}(E)\big]
	\;=\; \frac{\Gamma_{P_i}}{h}\!\int\! dE\,
	\Big(-\frac{\partial f_0}{\partial E}\Big)\,\alpha_{iL}(E),
	\\
	-\mathcal L^{(0)}_{ii} &= \sum_{\alpha\neq i} \mathcal L^{(0)}_{i\alpha}
	\;=\; \frac{\Gamma_{P_i}}{h}\!\int\! dE\,
	\Big(-\frac{\partial f_0}{\partial E}\Big)\big[\alpha_{iL}(E)+\alpha_{iR}(E)\big]
	\;+\; \mathcal O(\Gamma_{P_i}^2),
\end{align}
where transmissions between probes $i\!\to\! j$ are $\mathcal O(\Gamma_{P_i}\Gamma_{P_j})$ and can be neglected at this order.
Substituting these scalings into \eqref{eq:Schur_general} gives the probe contribution
\begin{equation}
	\Delta \mathcal L^{(0)}_{LR}
	\;=\;
	\sum_{i}
	\frac{\mathcal L^{(0)}_{L\,i}\,\mathcal L^{(0)}_{i\,R}}{-\mathcal L^{(0)}_{ii}}
	\;=\;
	\sum_{i} \Gamma_{P_i}\,
	\frac{
		\!\int dE\,(-f_0')\,\alpha_{iL}(E)\;\int dE\,(-f_0')\,\alpha_{iR}(E)
	}{
		\!\int dE\,(-f_0')\,\big[\alpha_{iL}(E)+\alpha_{iR}(E)\big]
	}
	\;+\; \mathcal O(\Gamma_{\rm P}^2).
	\label{eq:DeltaL0_linear}
\end{equation}

At a node $E_0$ the functions $\alpha_{i\alpha}(E)$ are smooth; over the thermal window they may be replaced by their values at $E_0$ up to $\mathcal O((k_BT)^2)$ corrections, yielding \eqref{eq:linear_floor_statement} and \eqref{eq:C_i_def}.

Equation~\eqref{eq:B_linear} shows that the incoherent background generated by $N$-probe connectivity is \emph{{linear}} 
in the total probe strength.  
This scaling follows directly from the D’Amato--Pastawski two-step formula,
$\Delta \mathcal T_{LR}=\sum_i \mathcal T_{L i}\,\mathcal T_{i R}/\sum_\gamma \mathcal T_{i\gamma}$:
the numerator scales as $\Gamma_{P_i}^2$ while the denominator scales as $\Gamma_{P_i}$, leaving a net $\Gamma_{P_i}$ dependence for each site.  
The geometry factor $\mathcal C_i(E_0)$ in Equation~\eqref{eq:C_i_def} encodes how efficiently orbital $i$ connects both leads; if a site is symmetry-dark to $L$ or $R$ (for example, due to a wavefunction node or another destructive symmetry), its $\mathcal C_i$ vanishes and that probe does not contribute.  
The resulting floor is therefore not universal but determined by geometry and the underlying quantum interference: distributed probes always open energy-independent bypasses, but~the magnitude of the background depends sensitively on which orbitals overlap both contacts.  
Beyond the wide-band approximation, the prefactor may develop curvature with $\Gamma_{\rm P}$, yet the linear law remains the leading contribution so long as the unprobed Green’s functions are smooth at $E_0$.  
Importantly, this linear floor sets the scale at which thermopower, figure of merit, and~efficiency saturate: once $B(\Gamma_{\rm P})$ exceeds the coherent transmission within the thermal window, the~probe-induced background dominates the response and erases distinctions between different nodal~orders.

\section{Effective Order $n_{\mathrm{eff}}(\Gamma_{\rm P})$ Under Weak Dephasing} 
\label{app:neff}

Near a nodal energy $E_0$ the coherent transmission has
\begin{equation}
	\mathcal T_{\rm coh}(E)
	= A_0\,|E{-}E_0|^{2n_0}
	+ \mathcal O\!\big(|E{-}E_0|^{2n_0+2}\big), 
	\qquad A_0>0,
\end{equation}
so all Onsager blocks inherit the power $2n_0+\nu$.  
Attaching a weak probe opens additional channels of \emph{lower} near-node order $n_{\rm low}$, with~amplitudes that scale with the total \linebreak  probe strength:
\begin{equation}
	\mathcal T_{\rm low}(E;\Gamma_{\rm P})
	= A_{\rm low}\,\Gamma_{\rm P}^{p}\,|E{-}E_0|^{2n_{\rm low}},
	\qquad n_{\rm low}\le n_0,\;\; p\simeq 1.
\end{equation}

Two cases capture the probe effects used in the main text:  
(i) a \emph{{dispersive}} 
lower-order bypass (single-site probe), where 
$n_{\rm low}=n_{\rm probe}=\max(b,c)$;  
(ii) an \emph{{energy-independent floor}} 
(all-site probes), where $n_{\rm low}=0$ and
\begin{equation}
	\mathcal T_{\rm low}\equiv B(\Gamma_{\rm P})
	\simeq C_{\rm geom}\,\Gamma_{\rm P}.
\end{equation}
To leading order in $\Gamma_{\rm P}$,
\begin{equation}
	\label{eq:app_Teff_general}
	\mathcal T_{\rm eff}(E;\Gamma_{\rm P})
	= B(\Gamma_{\rm P})\,|E{-}E_0|^{2n_{\rm low}}
	+ A_0\,|E{-}E_0|^{2n_0}
	+ \mathcal O\!\big(\Gamma_{\rm P}|E{-}E_0|^{2n_{\rm low}+2},\,\Gamma_{\rm P}^2\big).
\end{equation}
We define the instantaneous effective order by the log--log slope,
\begin{equation}
	n_{\rm eff}(E;\Gamma_{\rm P})
	\equiv \tfrac12\,\frac{d\ln \mathcal T_{\rm eff}}{d\ln|E{-}E_0|}.
\end{equation}
Inserting \eqref{eq:app_Teff_general} {gives} 
\vspace{-12pt}
\begin{align}
	n_{\rm eff}(E;\Gamma_{\rm P})
	&= n_{\rm low}
	+ (n_0-n_{\rm low})\,
	\frac{A_0\,|E{-}E_0|^{2(n_0-n_{\rm low})}}
	{B(\Gamma_{\rm P})+A_0\,|E{-}E_0|^{2(n_0-n_{\rm low})}} \nonumber\\[3pt]
	&= n_{\rm low}
	+ \frac{n_0-n_{\rm low}}{1+\big(\Gamma_{\rm P}/\Gamma_\ast(E)\big)^p},
	\label{eq:app_local_neff}
\end{align}
\begin{equation}
	\Gamma_\ast(E)
	= \Big(\tfrac{A_0}{A_{\rm low}}\Big)^{1/p}\,
	|E{-}E_0|^{\,2(n_0-n_{\rm low})/p}.
\end{equation}
Thus, the probe selects the dephased asymptote $n_\infty=n_{\rm low}$, while the coherent channel sets the clean asymptote $n_0$.

In practice we report a window-averaged slope (either over a fixed log-$k_B T$ range or weighted by the Fermi window).  
This average is well captured by the empirical Hill form used in the main text,
\begin{equation}
	\label{eq:app_neff_general}
	n_{\rm eff}(\Gamma_{\rm P})
	\simeq n_\infty
	+\frac{n_0-n_\infty}{\big(1+\Gamma_{\rm P}/\Gamma_c\big)^{\beta}},
\end{equation}
with
\begin{equation}
	n_\infty = 
	\begin{cases}
		\max(b,c), & \text{single-site bypass},\\[4pt]
		0, & \text{all-site floor},
	\end{cases}
	\qquad
	\Gamma_c \sim
	\Big(\tfrac{A_0}{A_{\rm low}}\Big)^{1/p}\,
	\Big\langle |E{-}E_0|^{\,2(n_0-n_\infty)}\Big\rangle_W^{1/p}.
\end{equation}

Here $\beta\approx p$ but also reflects details of the averaging~window.

The selection rule fixes the lower asymptote $n_\infty$.  
The crossover scale $\Gamma_c$ decreases with the order gap $\Delta n=n_0-n_\infty$, so higher-order supernodes ($n_0>1$) collapse at smaller $\Gamma_{\rm P}$ than quadratic nodes.  
The fitted exponent $\beta$ carries no universal meaning, but~$\Gamma_c$ provides a robust sensitivity measure: it is the value of $\Gamma_{\rm P}$ at which the effective order has decayed halfway between its coherent and dephased limits.

\section{The Influence of Node~Detuning}
\label{app:detune_nodes}
The order-selection rule described in the main text assumes that all relevant channels vanish at the same nodal energy.  
In practice, this condition need not hold.  
If the channels are \emph{{detuned}} 
—for example, $\mathcal T_{LR}$ vanishes at $E_a$, while $\mathcal T_{LP}$ and $\mathcal T_{PR}$ vanish at $E_b$ and $E_c$ with $E_{b,c}\neq E_a$—then near $E_a$ the probe-mediated pathway generically contributes finite~weight.  

For a VP this follows directly from
\begin{equation}
	\widetilde{\mathcal T}(E) \;=\; \mathcal T_{LR}(E) + \frac{\mathcal T_{LP}(E)\,\mathcal T_{PR}(E)}{\mathcal T_{LP}(E)+\mathcal T_{PR}(E)} ,
\end{equation}
so that a finite background
\begin{equation}
	\mathcal T_{\rm floor} \;\equiv\; \frac{\mathcal T_{LP}(E_a)\,\mathcal T_{PR}(E_a)}{\mathcal T_{LP}(E_a)+\mathcal T_{PR}(E_a)} \;>\;0
\end{equation}
emerges whenever the probe channels do not vanish at $E_a$.  
An analogous conclusion holds for a VTP at the level of Onsager moments: detuning injects a zeroth-order term into $\mathbf L^{\rm eff}_{LR}$.  

The consequences are straightforward.  
At asymptotically low $T$, transport is governed by $n_{\rm eff}=0$:
\begin{equation}
	G \;\simeq\; \frac{e^2}{h}\,\mathcal T_{\rm floor},\qquad 
	\kappa_e \;\simeq\; \frac{\pi^2 k_B^2 T}{3h}\,\mathcal T_{\rm floor},
\end{equation}
while the thermopower vanishes with temperature.  
If the intrinsic left--right channel scales as $|E{-}E_a|^{2a}$, then (up to ${\cal O}(1)$ prefactors)
\begin{equation}
	S_{\max}(T)\;\sim\;\frac{k_B}{e}\,\frac{(k_B T)^{2a}}{\mathcal T_{\rm floor}},\qquad 
	ZT_{\rm el}(T)\;\sim\;\frac{(k_B T)^{4a}}{\mathcal T_{\rm floor}^{\,2}}
	\;\xrightarrow[T\to 0]{}\;0.
\end{equation}

Thus, detuning does not protect supernode-enhanced thermopower; instead it erodes it by introducing a floor that dominates in the low-$T$ limit.  

Practically, apparent supernode scaling can still be observed over finite experimental ranges, provided the detuning-induced background is small on the thermal scale, i.e.,
\begin{equation}
	\mathcal T_{\rm floor}\ \ll\ A\,(k_B T)^{2a},
\end{equation}
with $A$ the prefactor of the coherent channel.  
But asymptotically the background wins, collapsing the enhancement.  
The suppression of electronic thermal conductance illustrates this point most clearly.  
At the nodal energy,
\begin{equation}
	\kappa(E_0)\;\propto\;(k_B T)^{2n_{\rm eff}+1},
\end{equation}
so that the BPDT junction with $n_{\rm eff}=1$ yields values many orders of magnitude below benzene ($n_{\rm eff}=0$).  
At room temperature ($k_B T\simeq0.026$\,eV), the~factor $(k_B T)^2\simeq7\times10^{-4}$\,eV$^2$ already enforces a dramatic reduction.  
Residual offsets arise from nodal amplitudes and from the $I_P^{(1)}=0$ constraint, which further restricts probe heat flow.  
Together with the dephasing analysis of the main text, these results confirm that temperature exposes the intrinsic exponent while dephasing renormalizes~it.

\bibliography{refs_clean}

\end{document}